\documentclass[preprint,prd,aps,superscriptaddress,11pt]{revtex4-1}

\usepackage{graphicx}
\usepackage{amsmath}
\usepackage{amssymb}
\usepackage{dcolumn}
\usepackage{bm}
\usepackage{slashed}
\usepackage{color}

\usepackage[normalem]{ulem}
\newcommand{\rs}[1]{{\color{red}\ifmmode\text{\sout{\ensuremath{#1}}}\else\sout{#1}\fi}}
\newcommand{\rx}[1]{{\color{red}\ifmmode\text{\xout{\ensuremath{#1}}}\else\sout{#1}\fi}}

\newcommand{\be}{\begin{equation}}
\newcommand{\ee}{\end{equation}}
\newcommand{\ba}{\begin{eqnarray}}
\newcommand{\ea}{\end{eqnarray}}
\newcommand{\no}{\nonumber \\}
\newcommand{\gsim}{\mathrel{\hbox{\rlap{\lower.55ex \hbox {$\sim$}}
                   \kern-.3em \raise.4ex \hbox{$>$}}}}
\newcommand{\lsim}{\mathrel{\hbox{\rlap{\lower.55ex \hbox {$\sim$}}
                   \kern-.3em \raise.4ex \hbox{$<$}}}}

\def\roughly#1{\mathrel{\raise.3ex\hbox{$#1$\kern-.75em%
\lower1ex\hbox{$\sim$}}}}
\def\lsim{\roughly<}
\def\gsim{\roughly>}

\def\({\left(}
\def\){\right)}
\def\[{\left[}
\def\]{\right]}
\def\<{\langle}
\def\>{\rangle}
\def\bv{{\bf v}}
\def\bp{{\bf p}}
\def\bE{{\bf E}}
\def\bB{{\bf B}}
\def\bO{{\bf\Omega}}
\def\bx{{\bf x}}
\def\tr{\text{tr}}
\def\tW{{\tilde W}}

\def\L{{\Lambda}}
\def\d{{\delta}}
\def\D{{\Delta}}
\def\o{{\omega}}

\def\e{{\epsilon}}

\def\b{{\beta}}
\def\c{{\chi}}

\def\p{{\pi}}

\def\m{{\mu}}
\def\n{{\nu}}

\def\s{{\sigma}}

\def\t{{\tau}}
\def\th{{\theta}}

\def\hb{{\hbar}}
\def\dg{{\dagger}}

\newcommand{\pd}{{\partial}}

\date{\today}

\begin{document}

\title{\bf Chiral kinetic theory from Landau level basis}

\author{Shu Lin}
\email{linshu8@mail.sysu.edu.cn}
\affiliation{School of Physics and Astronomy, Sun Yat-Sen University, Zhuhai 519082, China}
\author{Lixin Yang}
\email{yanglx5@mail2.sysu.edu.cn}
\affiliation{School of Physics and Astronomy, Sun Yat-Sen University, Zhuhai 519082, China}

\begin{abstract}

We derive a chiral kinetic theory with Landau level basis, which is valid for slow-varying magnetic field with arbitrary magnitude. We apply the new chiral kinetic theory to calculate the electric conductivity transverse to the magnetic field in a magnetized QED and QCD plasma. Under the lowest Landau level approximation and relaxation time approximation, we find the transverse conductivity is inversely proportional to the relaxation time. We also obtain a frequency-dependent transverse conductivity in response to a time-dependent electric field. We find a high frequency enhancement in this conductivity.

\end{abstract}

\maketitle


\newpage

\section{Introduction}

The physics of chiral fermions have attracted much attention recently. Through chiral anomaly, system of chiral fermions show intriguing non-dissipative transport phenomena such as chiral magnetic effect \cite{Vilenkin:1980fu,Kharzeev:2004ey,Kharzeev:2007tn,Fukushima:2008xe,Son:2009tf,Neiman:2010zi} and chiral vortical effect \cite{Vilenkin:1980zv,Erdmenger:2008rm,Banerjee:2008th,Son:2009tf,Neiman:2010zi,Landsteiner:2011cp}. There has been growing interest in the verification of these anomalous transports in both quark-gluon plasma \cite{Adamczyk:2014mzf,Abelev:2009ac,Abelev:2012pa,Sirunyan:2017quh} and Dirac/Weyl semimetal \cite{Li:2014bha,Gooth:2017mbd}. Chiral kinetic theory (CKT), as a theoretical tool to describe dynamics of chiral fermions has been undergoing rapid development over the past few years. It has been derived from different approaches \cite{Son:2012wh,Son:2012zy,Stephanov:2012ki,Gao:2012ix,Pu:2010as,Chen:2012ca,Hidaka:2016yjf,Manuel:2013zaa,Manuel:2014dza,Wu:2016dam,Mueller:2017arw,Mueller:2017lzw,Huang:2018wdl,Gao:2018wmr,Carignano:2018gqt,Lin:2019ytz,Carignano:2019zsh,Liu:2018xip,Dayi:2018xdy,Weickgenannt:2019dks,Gao:2019znl,Hattori:2019ahi,Wang:2019moi}

Formally, the CKT can be understood as an expansion in the reduced Planck constant $\hb$ \cite{Hidaka:2016yjf,Huang:2018wdl,Gao:2018wmr}. Since $\hb$ is always accompanied by external electromagnetic field and spacetime gradient, we may also view the expansion in weak field and spacetime disturbance. The lowest order in the expansion simply gives the Boltzmann equation, which describes dynamics of classical particle.
\begin{align}
  \(\pd_t+\bv\cdot{\nabla_\bx}+Q(\bE+\bv\times\bB)\cdot{\nabla_\bp}\)f(X,\bp)=C[f], 
\end{align}
with $Q$ being charge of the fermion.
At next to leading order, the quantum effect of spin of chiral fermions is manifest.
\begin{align}
  &\big[\(1+\hb Q{\bO}\cdot\bB\)\pd_t+\(\bv+\hb Q\bE\times\bO+\hb Q{\(\bv\cdot\bO\)}\bB\)\cdot\nabla_\bx \no
  &+\(Q\bE+Q\bv\times\bB+{\hb}Q^2(\bE\cdot\bB)\bO\)\cdot\nabla_\bp\big]
  f(X,\bp)=C[f].
\end{align}
It leads to a quasi-particle with magnetic moment and Berry curvature $\bO=\pm\frac{\hat \bp}{2\bp^2}$ with plus/minus sign for right/left handed fermions respectively. The appearance of $\bO$ is essential for correct description of chiral anomaly. However the situation becomes quite cumbersome at next-to-next-to leading order \cite{Gao:2018wmr}, in which case not only the quasi-particle picture is lost, but also the resulting CKT becomes more singular in the IR. This is not surprising as in the regime $\bp^2\sim \hb B$, Landau quantization becomes relevant invalidating quasi-particle picture.

The difficulty of the above expansion arises due to an inappropriate choice of basis, for which the lowest order is given by free particle with the dispersion $p_0^2=\bp^2$. With weak magnetic field perturbation, the basis has to be shifted: to first order in $B$ the shift leads to modified dispersion $p_0^2=\bp^2-{\hb Q\bB\cdot\bp/p_0}$. In fact, we may choose a different basis from the beginning. A useful basis is the Landau level basis, which is suitable for a constant magnetic field $\hb B\sim \bp^2$. There has been early attempt on formulating CKT with lowest Landau level (LLL) \cite{Hattori:2016lqx}, which is valid for strong magnetic field and the motion of LLL is restricted to be parallel to the magnetic field. In this paper, we derive a CKT with full Landau level basis more systematically using Wigner function approach. The resulting CKT is valid to all order in $\hb B$, but only to first order in perturbation other than the background magnetic field. Moreover, the motion of Landau levels is not restricted.

The paper is organized as follows: in Section \eqref{sec:ckt}, we derive a CKT with Landau level basis using Wigner function approach. It is formally the same as CKT with free particle basis but with a different power counting. We will show to the lowest order in perturbation it is satisfied by Landau levels. As an application of the CKT, we study transverse conductivity of a plasma consisting of chiral fermions subject to external magnetic field in Section \eqref{sec:sigma}. For simplicity, we employ relaxation time approximation for the collision term and use LLL approximation, which is valid for strong magnetic field. We conclude and discuss future works in Section \eqref{sec:summary}.

For convention, we absorb electric charge $e$ into the gauge field and take mostly negative Minkowski metric. We also set $\hb=1$ in the rest of this paper.

\section{Chiral kinetic theory with Landau level basis}\label{sec:ckt}

To be specific, we start with a system of right-handed chiral fermions with positive charge $Q=1$ in unit of electric charge $e$. The Wigner function corresponding to this system is given by
\begin{align}\label{wigner}
  {\tW(X,P)}=\int\frac{d^4X'}{(2\pi)^4}e^{-iP\cdot X'}\<\psi(X-\frac{1}{2}X')U(X+\frac{1}{2}X',X-\frac{1}{2}X')\psi^\dg(X+\frac{1}{2}X')\>.
\end{align}
Here $\psi$ is Weyl spinor and ${\tW(X,P)}$ is a $2\times2$ matrix in Dirac space. $X^\m=(t,x,y,z)$ and $P^\m=(p_0,p_x,p_y,p_z)$ will be interpreted as coordinate and momentum of the resulting CKT {with $P\cdot X=P_\m X^\m$}\footnote{Note that $P_\m=(p_0,-p_i)$ where we let Roman indices such as $i,j$ run over $1,2,3$ or $x,y,z$.}.
The following gauge link $U(X+\frac{1}{2}X',X-\frac{1}{2}X')$ is inserted to ensure gauge invariance of ${\tW(X,P)}$
\begin{align}\label{link}
  U(X+\frac{1}{2}X',X-\frac{1}{2}X')=\mathrm{exp}\[i\int_{X-\frac{1}{2}X'}^{X+\frac{1}{2}X'}dR^\m\(A_\m(R)+a_\m(R)\)\].
\end{align}
The gauge link is path-dependent. We take the path to be a straight line.
Note that we have splitted the gauge potential into $A_\m$ and $a_\m$ corresponding to background magnetic field and perturbation respectively. We denote $F_{\m\n}=\pd_\m A_\n-\pd_\n A_\m$ and $f_{\m\n}=\pd_\m a_\n-\pd_\n a_\m$ as field strength for constant magnetic field and perturbation. Below we keep to all order in $B$ and only first order in the perturbation.

To derive the kinetic equations, we start with the following equation of motion (EOM):
\begin{align}\label{eom_xy}
  {\slashed D}_Z\<\psi(Z)\psi(Y)^\dg\>=0,\quad \<\psi(Z)\psi(Y)^\dg\>{\slashed D}_Y^\dg=0,
\end{align}
with the covariant derivatives defined as
\begin{align}
  &  {\slashed D}_Z={\slashed \pd}_Z+i{\slashed A}(Z)+i{\slashed a}(Z), \no
  &  {\slashed D}_Y^\dg=\overleftarrow{{\slashed \pd}}_Y-i{\slashed A}(Y)-i{\slashed a}(Y).
\end{align}
For right-handed fermion, the slash is given by ${\slashed a}=\s^\m a_\m$. Denoting $\<\psi(Z)\psi(Y)^\dg\>={\bar W(Z,Y)}$ and setting $Z=X-\frac{1}{2}X'$, $Y=X+\frac{1}{2}X'$, we can relate ${\bar W(Z,Y)}\equiv W(Z,Y)U(Z,Y)$ to Wigner function as
\begin{align}
  &  {\tW(X,P)}=\int\frac{d^4X'}{(2\pi)^4}e^{iP\cdot (Z-Y)}W(Z,Y) \no
  &=\int\frac{d^4X'}{(2\pi)^4}e^{iP\cdot (Z-Y)}{\bar W(Z,Y)}U(Y,Z).
\end{align}
Using \eqref{eom_xy}, we can derive EOM satisfied by ${\tW(X,P)}$. To proceed, we work out explicit expression for $U(Y,Z)$. We need the following gradient expansion of the background and perturbation.
\begin{align}\label{Aa_exp}
&  A_\m(X+\D X)=A_\m(X)+\(\D X{\cdot\pd_X}\)A_\m(X), \no
&  a_\m(X+\D X)=a_\m(X)+\(\D X{\cdot\pd_X}\)a_\m(X)+O\((\pd_X)^2\).
\end{align}
Note that for the constant background magnetic field, the background gauge potential is linear in coordinate so the expansion truncates at first order. For the perturbation, we keep only first order in gradient. It follows that $U(Y,Z)=e^{-iX'\cdot[A(X)+a(X)]}+O\((\pd_X)^2\)$. Using this we can rewrite the EOM \eqref{eom_xy} as
\begin{align}\label{eom_U}
  {\slashed D}_Z\(W(Z,Y)e^{iX'\cdot\[A(X)+a(X)\]}\)=\(W(Z,Y)e^{iX'\cdot\[A(X)+a(X)\]}\){\slashed D}_Y^\dg=0.
\end{align}
We can exchange covariant derivative with gauge link using the following identities, which hold up to first order in gradient
\begin{align}
  &  D_{Z\m}\(W(Z,Y)e^{iX'\cdot\[A(X)+a(X)\]}\)=e^{iX'\cdot A(X)}\(\frac{1}{2}\pd_{X\m}-\pd_{X'\m}+\frac{i}{2}X'^\n(F_{\m\n}+f_{\m\n})\)W(Z,Y),\no
  &  \(W(Z,Y)e^{iX'\cdot\[A(X)+a(X)\]}\)D_{Y\m}^\dg=W(Z,Y)\(\frac{1}{2}\overleftarrow{\pd}_{X\m}+\overleftarrow{\pd}_{X'\m}+\frac{i}{2}X'^\n(F_{\m\n}+f_{\m\n})\)e^{iX'\cdot A(X)},
\end{align}
with $\pd_{X\m}\equiv\pd/\pd X^\m$.
Note that $W$ don't commute with $\s^\m$. We obtain the following EOM for $W(Z,Y)$
\begin{align}\label{eom_W}
  &  \(\frac{1}{2}\pd_{X\m}-\pd_{X'\m}+\frac{i}{2}X'{}^\n(F_{\m\n}+f_{\m\n})\)\s^\m W(Z,Y)=0, \no
  &  \(\frac{1}{2}\pd_{X\m}+\pd_{X'\m}+\frac{i}{2}X'{}^\n(F_{\m\n}+f_{\m\n})\) W(Z,Y)\s^\m=0.
\end{align}
Fourier transforming the above equations with $\int\frac{d^4X'}{(2\pi)^4}e^{-iP\cdot X'}$, we arrive at EOM for Wigner function ${\tW}(X,P)$:
\begin{align}\label{eom_wigner}
  &(\frac{1}{2}\D_\m-iP_\m)\s^\m{\tW}(X,P)=0, \no
  &(\frac{1}{2}\D_\m+iP_\m){\tW}(X,P)\s^\m=0,
\end{align}
where we have defined $\D_\m\equiv\pd_{X\m}-\frac{\pd}{\pd P_\n}\(F_{\m\n}+f_{\m\n}\)$\footnote{Explicitly, one has $\D_\m=(\D_0,\D_i)=(\pd_{X0}+\(F_{0k}+f_{0k}\)\frac{\pd}{\pd p_k},\,\pd_{Xi}-\(F_{i0}+f_{i0}\)\frac{\pd}{\pd p_0}+\(F_{ij}+f_{ij}\)\frac{\pd}{\pd p_j})$.}.
We can further decompose the matrix equations into components. Note that $W(Z,Y)=W(Y,Z)^\dg$. It follows that ${\tW}(X,P)$ is hermitian, so that we can decompose it in the basis of $\s^\m$:
\begin{align}\label{decomp}
  {\tW}(X,P)=\frac{1}{2}\(F(X,P)\cdot{\bf1}+j_i(X,P)\s^i\).
\end{align}
$F$ and $j_i$ are related to charge density $J_0$ and current density $J_i$ by momenta integration:
\begin{align}\label{Fj}
  &J_0(X)=\int d^4P\;\tr{\tW}(X,P)=\int d^4P\;F(X,P),\no
  &J_i(X)=\int d^4P\;\tr{\tW}(X,P)\s^i=\int d^4P\;j_i(X,P).
\end{align}
We can then project \eqref{eom_wigner} into the basis $\s^\m$ to obtain
\begin{align}\label{eom_comp}
  \D_0F+\D_ij_i=0, \no
  \D_0j_i+\D_iF-2\e^{ijk}p_jj_k=0, \no
  p_0F-p_ij_i=0, \no
  -p_0j_i+p_iF+\frac{1}{2}\e^{ijk}\D_jj_k=0.
\end{align}
The first two lines of \eqref{eom_comp} can be viewed as transport equations for $F$ and $j_i$ respectively. The last two lines of \eqref{eom_comp} are constraint equations. They are formally the same as CKT with free particle basis, but with a different power counting: we regard $\pd_{X\m}$ and $\frac{\pd}{\pd P_\n}f_{\m\n}$ as first order perturbation and $\frac{\pd}{\pd P_\n}F_{\m\n}$ and $P^\m$ as zeroth order. Consequently, the zeroth order solution of Wigner function is not given by superposition of free particle contribution, but rather superposition of Landau level contribution. In appendix A we verify explicitly that individual Landau level contribution satisfies \eqref{eom_comp}.

Let's comment on the regime of applicability of the CKT. To be specific, we point the magnetic field in the $z$ direction. The explicit LL solutions imply that $B\sim p_T^2$, with $p_T=\sqrt{p_x^2+p_y^2}$ being the modulus of the transverse momentum and the longitudinal momentum ${p_z}$ is unrestricted. Furthermore, the perturbation should be much smaller than the background, thus ${f_{\m\n}}\ll B$ and $\pd_X\ll \sqrt{B}$.

Before closing this section, we show the CKT \eqref{eom_comp} reduces to the equations written down for studying longitudinal conductivity in strong magnetic field limit in \cite{Hattori:2016lqx}. To this end, we turn on an electric field parallel to the magnetic field. This would induce longitudinal motion of LLL. Since the longitudinal motion is classical. We can simply use the following ansatz for the Wigner function components:
\begin{align}\label{comp_lll}
  F=j_3=\frac{{2}}{(2\pi)^3}\mathrm{exp}\(-\frac{p_T^2}{B}\)\Big(f_+(X,p_0)\d(p_0-{E_{p_z}})+f_-(X,p_0)\d(p_0+E_{p_z})\Big),
\end{align}
where $f_{\pm}$ are distribution functions of positive energy and negative energy LLL states. It is easy to see the above ansatz automatically satisfies constraint equations by noting massless $E_{p_z}=|p_z|$ and $p_z>0/p_z<0$ for positive/negative energy LLL states respectively. Plugging this into the dynamical equations, we readily obtain
\begin{align}\label{ckt_lll}
  \(\frac{\pd }{\pd t}+\frac{\pd}{\pd z}+E_z\Big(\frac{\pd}{\pd p_0}{+\frac{\pd}{\pd p_z}}\Big)\)f_\pm(X,p_0)\d(p_0-{p_z})=0.
\end{align}
In writing down \eqref{ckt_lll} we adopt a non-standard point of view: instead of treating the system as a collection of positively and negatively charged LL states both with positive energies, we equivalently regard the system as a collection of positively charged LL states with positive and negative energies. 
Since the LLL states are all positively charged, we have $\dot{z}=\frac{p_z}{p_0}=1$, $E_z=\dot{p}_z$.
Noting that the overall operator passes through $\d(p_0-p_z)$, we simply obtain
\begin{align}\label{ckt_lll_simp}
  \(\frac{\pd }{\pd t}+\frac{\pd}{\pd z}+E_z\frac{\pd}{\pd p_0}\)f_\pm(X,p_0)=0.
\end{align}
The advantage of the above point of view treats states with all energies on the equal footing, so we expect the distribution functions to be continuous at $p_0=0$: $f_+(p_0=0)=f_-(p_0=0)$. Now we switch to standard picture consisting of positively and negatively charged states, all with positive energies. We denote their corresponding distribution functions with ${\bar f}_+$ and ${\bar f}_-$ respectively. The distribution functions in the two pictures are related by
\begin{align}\label{pic_trans}
  f_+(X,p_0)={\bar f}_+(X,|p_0|),\quad f_-(X,p_0)=1-{\bar f}_-(X,|p_0|),
\end{align}
Note that the Wigner function is normal ordered for positive and negative energy states. Converting the negative energy states to negatively charged states amounts to exchanging the corresponding creation and annihilation operators. The resulting anti-normal ordered term can be made normal ordered by using anti-commutation relation, which introduces the additional $1$ in \eqref{pic_trans}, which is state independent, as has been emphasized recently in \cite{Sheng:2018jwf,Gao:2019zhk}. Then $f_+(p_0=0)=f_-(p_0=0)$ implies ${\bar f}_+(X,|p_0|=0)+{\bar f}_-(X,|p_0|=0)=1$.
Note that momentum flips sign for negatively charged state, thus $p_0$-derivative term should flip sign accordingly. We readily identify \eqref{ckt_lll} with the collisionless limit of the kinetic equation used by \cite{Hattori:2016lqx}.
Note that \eqref{eom_comp} doesn't contain a collisional term because the system we consider is chiral fermions minimally coupled to external magnetic field. We will introduce a collisional term in the next section.

We can reproduce chiral anomaly equation by using \eqref{ckt_lll} as follows:
\begin{align}
  &\pd_\m J^\m=\int d^4P\frac{{2}}{(2\pi)^3}\mathrm{exp}\(-\frac{p_T^2}{B}\)\(\frac{\pd }{\pd t}+\frac{\pd}{\pd z}\)\({\bar f}_+(X,|p_0|)\th(p_0)-{\bar f}_-(X,|p_0|)\th(-p_0)\)\d(p_0-p_z)\no
  &=-\int d^4P\frac{{2}}{(2\pi)^3}\mathrm{exp}\(-\frac{p_T^2}{B}\)\,E_z\Big(\frac{\pd}{\pd p_0}{+\frac{\pd}{\pd p_z}}\Big)\({\bar f}_+(X,|p_0|)\th(p_0)-{\bar f}_-(X,|p_0|)\th(-p_0)\)\d(p_0-p_z)\no
&=-\frac{E_zB}{(2\p)^2}\int dp_0\,\frac{\pd}{\pd p_0}\({\bar f}_+(X,|p_0|)\th(p_0)-{\bar f}_-(X,|p_0|)\th(-p_0)\)\no
&=-\frac{E_zB}{(2\p)^2}\(-{\bar f}_+(X,p_0=0)-{\bar f}_-(X,p_0=0)\)=\frac{E_zB}{(2\p)^2}.
\end{align}
In the last equality, we have used $f_\pm(p_0\to\infty)=0$. Note that we have a system of right-handed fermions. The above current corresponds to right-handed current $J_R$. Contribution from left-handed current would give an opposite contribution. Combining both contributions, we get $\pd_\m J^\m=0$ and $\pd_\m J_5^\m=\frac{\bE\cdot \bB}{2\p^2}$. The above derivation clearly shows the origin of chiral anomaly is from the term $1$ in \eqref{pic_trans} which is identified as vacuum contribution. Therefore it's consistent with the vacuum origin of chiral anomaly.
Since only the lowest Landau level contributes to $j_z$, we can reproduce chiral magnetic effect and chiral separation effect from \eqref{comp_lll} with equilibrium distribution. This has been done in \cite{Sheng:2017lfu}.


\section{Application: transverse conductivity}\label{sec:sigma}

In this section, we study transverse conductivity of the system. As we shall see, the simple ansatz used in the longitudinal case is inadequate. Physically transverse motion of LL states is quantum. If a transverse electric field is applied, the LL basis becomes incomplete. Nevertheless, we can solve the system to first order to the electric field perturbation to obtain the deviation from LL basis. Similar phenomenon occurs in the perturbation by magnetic field with free particle basis, which leads to magnetic moment modified dispersion relation and Berry curvature modified dynamical equations. We will work it out in detail.

\subsection{Collisional term}

As we mentioned in the previous section, since conductivity is dissipative transport coefficients, we need to introduce a collisional term. For simplicity, we use relaxation time approximation for the collisional term, which leads to the following modified dynamical equations
\begin{align}\label{rta}
    \D_0F+\D_ij_i=-\frac{\d F}{\t}, \no
  \D_0j_i+\D_iF-2\e^{ijk}p_jj_k=-\frac{\d j_i}{\t},
\end{align}
with $\t$ being the relaxation time. $\d F$ and $\d j_i$ measures deviation of equilibrium value of Wigner function components in rest frame of the plasma system. This is also the same frame with a constant background magnetic field. We will solve \eqref{rta} in the presence of transverse electric field. Keen readers may have noticed that \eqref{rta} is only a subset of the full equations. We now clarify the reason for keeping the dynamical equations only.

We first note that the CKT \eqref{eom_comp} is an over-determined set of equations, with four fields but eight equations including four constraint equations and four dynamical equations. The constraint equations don't contain time derivatives while the dynamical equations do. From the point of view of the evolution of the system, the constraint equations continue to hold as the system evolves with dynamical equations once it is satisfied by the initial condition. We show this for the case without collisional term explicitly in Appendix B.

The situation changes as we turn on {\it transverse} electric field. As we remarked already, unlike parallel electric field, transverse electric field induces quantum motion of LL states, driving the system off the Hilbert space expanded by the LL states. This is not something new to us. Similar things occur when we turn on magnetic field to free particle states, in which the Hilbert space is modified from the one expanded by free particle states to the counterpart by quasi-particle states. In this case, the external magnetic field modifies the constraint equations as well, leading to modified dispersion. Similarly, we expect the addition of collisional term also modify the constraint equations in such a way that over-determined system is self-consistent. In principle, both the collisional term and modification to constraint equations should follow from the same self-energy of the chiral fermion \cite{Hidaka:2016yjf}. However, since we use an ad-hoc relaxation type collisional term, it is not obvious how to modify the constraint equations. We shall not attempt it here but simply solve the dynamical equations \eqref{rta}.

\subsection{Transverse conductivity}

We are interested in calculating transverse conductivity of QED plasma at constant chemical potential and temperature. To simplify the calculation, we consider the limit of strong magnetic field, for which we can use the LLL approximation \eqref{comp_lll} for the background Wigner function components.
We first turn on a constant $E_i=f^{i0}$ transverse to the magnetic field $B_i=-\frac{1}{2}\e^{ijk}F_{jk}$, then $\D_\m$ reduce to $-\(F_{\m\n}+f_{\m\n}\)\pd/\pd P_\n$ since $\pd_{X\m}=0$. For the calculation of conductivity, it is sufficient to keep responses of Wigner function components up to linear order in $E$. We use $\d F$ and $\d j_i$ to denote these responses.
To linear order in $E$, the dynamical equations \eqref{rta} become
\begin{align}\label{pert_CKT}
  &-\frac{2E_ip_i}{B}F+B\e^{ij}\frac{\pd}{\pd p_i}\d j_j=-\frac{\d F}{\t}, \no
  &-\frac{2E_ip_i}{B}F-2\e^{ij}p_i\d j_j=-\frac{\d j_3}{\t}, \no
  &E_i\frac{\pd F}{\pd p_0}-B\e^{ij}\frac{\pd}{\pd p_j}\d F-2\e^{ij}p_j\d j_3+2\e^{ij}\d j_jp_z=-\frac{\d j_i}{\t}.
\end{align}
We have used the property $j_3=F$ and $\frac{\pd F}{\pd p_i}=-\frac{2p_i}{B}F$. From now on, we reserve the indices $i,j$ etc. for transverse coordinates only. Eliminating $\d F$ and $\d j_3$ from \eqref{pert_CKT}, we obtain the following equation for $\d j_i$:
\begin{align}\label{dji_CKT}
  -\frac{\d j_i}{\t}&-\t B^2\e^{ij}\e^{kl}\frac{\pd^2}{\pd p_j\pd p_k}{\d j_l}+4\t \e^{ij}\e^{kl}p_jp_k\d j_l{-}2\e^{ij}\d j_jp_z 
  =E_i\frac{\pd}{\pd p_0}F-2\t\e^{ij}E_jF.
\end{align}
Note that the RHS $\propto\mathrm{exp}\(-\frac{p_T^2}{B}\)$. It follows that we can take the ansatz
\begin{align}\label{ansatz}
  \d j_i=E_i\d j_{\parallel}+\e^{ij}E_j\d j_\perp,
\end{align}
with $j_\parallel, j_\perp\propto\mathrm{exp}\(-\frac{p_T^2}{B}\)$. This converts \eqref{dji_CKT} into an algebraic equation
\begin{align}\label{alg}
  \begin{pmatrix}
    -2\t B-\frac{1}{\t}& 2p_z\\
    -2p_z& -2\t B-\frac{1}{\t}
  \end{pmatrix}\begin{pmatrix}
    \d j_\parallel\\
    \d j_\perp
    \end{pmatrix}=
  \begin{pmatrix}
    \frac{\pd F}{\pd p_0}\\
    -2\t F
  \end{pmatrix}.
\end{align}
then, the solutions are given by
\begin{align}\label{j1j2}
&\d j_\parallel=\frac{4p_z\t F-\(2B\t+\frac{1}{\t}\)\frac{\pd F}{\pd p_0}}{\(2B\t+\frac{1}{\t}\)^2+4p_z^2},\no
&\d j_\perp=\frac{2p_z\frac{\pd F}{\pd p_0}+2\(2B\t+\frac{1}{\t}\)\t F}{\(2B\t+\frac{1}{\t}\)^2+4p_z^2}.
\end{align}
The fact that the perturbed components $\d j_i$ deviate from the space formed by the LLL basis is clearly visible in the structure $\frac{\pd F}{\pd p_0}$, generating terms $\propto\d'(p_0-p_z)$, which are absent in the LLL basis.
Obviously, $\d j_i$ are even functions of $p_i$. From \eqref{pert_CKT}, the resulting solutions for $\d F$ and $\d j_3$ are odd functions of $\bp_T$, and therefore vanish upon integration over momenta. This means a transverse perturbation $E$ does not induce electric charge density and longitudinal current at this order. The transverse currents given by \eqref{Fj} are nontrivial. After the integration over momenta detailed in Appendix C, the electric current induced by a perturbative $\bE$ transverse to background $\bB$ 
can be written as
\begin{align}
e\d{\bf J}=\s_{\perp}\bE+\s_H\frac{\bB\times\bE}{B},
\end{align}
where
\begin{align}\label{sig_perp}
\s_{\perp}&=\frac{e^3B}{(2\p)^2}\[\int_{0}^{\infty}\(\frac{1}{e^{\b(p_z-\m)}+1}+\frac{1}{e^{\b(p_z+\m)}+1}\)\frac{4\t p_z}{\(2eB\t+\frac{1}{\t}\)^2+4p_z^2}dp_z+\frac{1}{2eB\t+\frac{1}{\t}}\]\no
&\xrightarrow{eB\gg p_z^2}\frac{e^2}{8\p^2\t}\[1+\frac{1}{eB}\(\m^2+\frac{\p^2T^2}{3}-\frac{1}{2\t^2}\)\]
\end{align}
\begin{align}\label{sig_H}
\s_H&=-\frac{2e^3B}{(2\p)^2}\int_{0}^{\infty}\(\frac{1}{e^{\b(p_z-\m)}+1}-\frac{1}{e^{\b(p_z+\m)}+1}\)\frac{2eB\t^2+1}{\(2eB\t+\frac{1}{\t}\)^2+4p_z^2}dp_z\no
&\xrightarrow{eB\gg p_z^2}-\frac{e^2\m}{4\p^2}.
\end{align}
Here we've restored electric charge unit $e$ in $\s_{\perp}$ and $\s_H$ and take the strong magnetic field limit in the end. For $\s_\perp$, we keep the leading two terms in $1/B$ expansion, while for $\s_H$, we keep only the leading term. Note that these results of conductivity are for magnetized plasma consisting of right-handed fermions only. To include the contribution from left handed fermions, we need to multiply it by a factor of $2$.

It is instructive to compare the results of conductivities of QED plasma magnetized by weak magnetic field in \cite{Gorbar:2016qfh,Chen:2016xtg,Hidaka:2017auj}. In the latter case, the ordinary conductivity is isotropic $\s\propto \t_0\chi$, with $\t_0$ being the relaxation time, which is in general different from $\t$ in the strong magnetic field case. The Hall conductivity depends on the relaxation time $\t_0$ as $\s_H\propto \m\t_0^2B$ thus is dissipative. The situation is quite different in the case with strong magnetic field. The ordinary conductivity is highly anisotropic. The longitudinal conductivity has been estimated in \cite{Hattori:2016lqx}, see also \cite{Hattori:2016cnt} as $\s_\parallel\sim \t_\parallel B^2/\chi\sim \t_\parallel B$, where the susceptibility $\chi\sim B$ in the strong $B$ limit. It has the same linear dependence on the relaxation time as the weak magnetic field case. On the contrary, the transverse conductivity we obtain scales as $\s_\perp\sim 1/\t_\perp$.
Note that the relaxation times $\t_\parallel$ and $\t_\perp$ responsible for longitudinal and transverse conductivities are in general different. For $\t_\parallel$, field theoretic analysis in \cite{Hattori:2016lqx} gives $\t_\parallel\sim O(B^0)$, which is consistent with recent lattice results on conductivities in magnetized quark-gluon plasma (QGP) \cite{Astrakhantsev:2019zkr}. For $\t_\perp$, a field theoretic analysis is still missing. A naive comparison with lattice results \cite{Astrakhantsev:2019zkr} seems to suggest a nontrivial dependence on $B$ for $\t_\perp$. Calculating $\t_\perp$ from first principle is beyond the scope of this paper, we will treat $\t_\perp$ as a phenomenological parameter, assumed to be independent of $B$ and denoted by $\t$ for simplicity.
The inverse proportionality $\s_\perp\sim1/\t$ may be counter-intuitive. It seems to imply a system with very strong dissipation $\t\to0$ would have large transverse conductivity. This is not true. In fact \eqref{alg} suggests an effective relaxation time $\frac{1}{2\t B+1/\t}$, which arises from an interplay of $\d F$ and $\d j_i$. The scaling $\s_\perp\sim 1/\t$ is valid in the limit $\t\gg B^{-1/2}$ only.
Finally, the Hall conductivity we obtain $\s_H\sim \m$. It is independent of $\t$, thus is non-dissipative.

The above results can be easily extended to the case with time-dependence. Let us consider the electric field perturbation $E\;\propto\;e^{-i\o t}$. From \eqref{rta}, it is clear $\pd_t$ and $\frac{1}{\t}$ always appear in the combination $\pd_t+\frac{1}{\t}$. It follows that we can obtain the solution in the time-dependent case with the following replacement in \eqref{sig_perp} and \eqref{sig_H} $\t\to\t_\o\equiv\frac{\t}{1-i\o\t}$. $\t_\o$ can be viewed as a complex relaxation time. An interesting limit is $\o\t\gg1$. This would give a small purely imaginary relaxation time $\t_\o\to\frac{i}{\o}$. Plugging this into \eqref{sig_perp}, we obtain a large purely imaginary conductivity from the leading term. This corresponds to an enhancement in the magnitude of $\s_\perp$ and a phase delay in the response to transverse $E$ field. Similar enhancement is also observed in \cite{Li:2018ufq}.

To apply the above results to magnetized QGP, we generalize the results to quantum chromodynamics (QCD) with $N_c$ colors and $N_f$ flavors. This amounts to summing over colors and flavors, as well as both left-handed and right-handed quark contributions, giving an overall factor $2N_c\sum_fQ_f^2$ to \eqref{sig_perp}:
\begin{align}\label{sig_qgp}
  \s_{\perp}&=2N_c\sum_f\frac{Q_f^2e^3B}{(2\p)^2}\Bigg[\int_{0}^{\infty}\(\frac{1}{e^{\b(p_z-\m_q)}+1}+\frac{1}{e^{\b(p_z+\m_q)}+1}\)\frac{4\t_\o p_z}{\(2|Q_f|eB\t_\o+\frac{1}{\t_\o}\)^2+4p_z^2}dp_z \no
  &+\frac{1}{2|Q_f|eB\t_\o+\frac{1}{\t_\o}}\Bigg]
\xrightarrow{eB\gg p_z^2}2N_c\sum_f\frac{Q_f^2e^2}{8\p^2\t_\o}\[1+\frac{1}{|Q_f|eB}\(\m_q^2+\frac{\p^2T^2}{3}-\frac{1}{2\t_\o^2}\)\].
\end{align}
Here we use $\m_q$ for quark chemical potential as counterpart of $\m$ in the QED case. Since QGP produced in heavy ion collisions is charge neutral, the Hall conductivity is negligible.

\begin{figure}
  \includegraphics[width=0.6\textwidth]{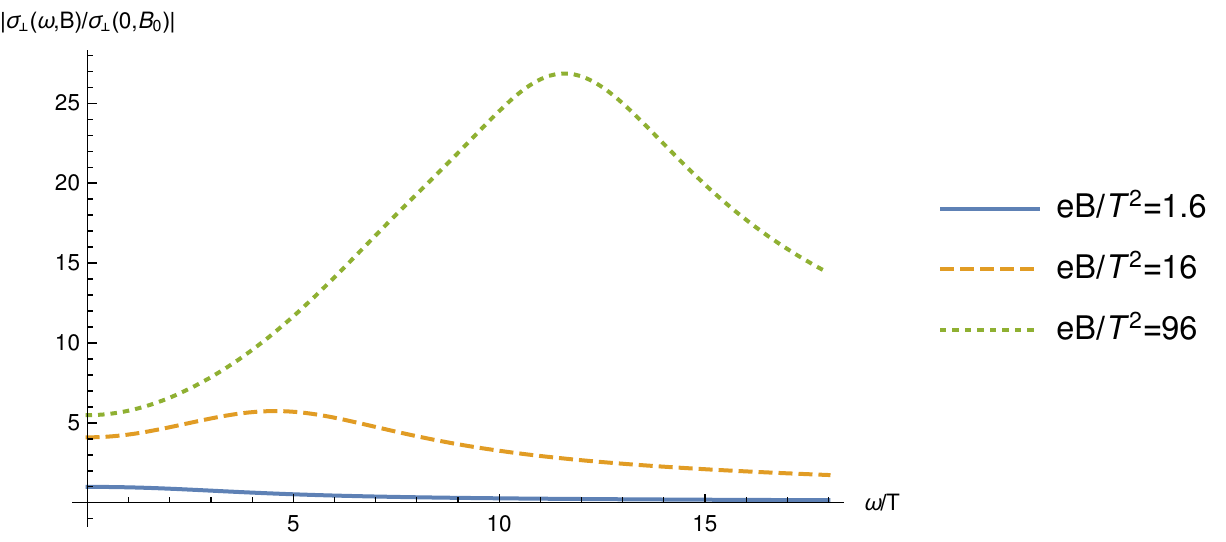}
  \caption{$\o$-dependence of the transverse conductivity for three different values of $B$ with $\t\simeq0.17\text{fm}$, i.e., $\t T\simeq0.3$. We take $\t_B\ge0.2\text{fm}$ corresponding to $\o/T\leqq 17.9$ and normalize the conductivity by its value at $eB_0/T^2=1.6$ corresponding to $eB_0\simeq 10m_\pi^2$. The dotted and dashed lines indicate an enhancement of conductivity when $\o\lesssim\sqrt{eB}$. The LLL approximation breaks down and the conductivity drops at higher $\o$ where one needs contribution from higher Landau Levels. For a HIC relevant $eB_0\simeq 10m_\pi^2$, the solid line shows no enhancement of conductivity which also calls for contributions of HLL.}
  \label{fig:o_dpd1}
\end{figure}

\begin{figure}
  \includegraphics[width=0.6\textwidth]{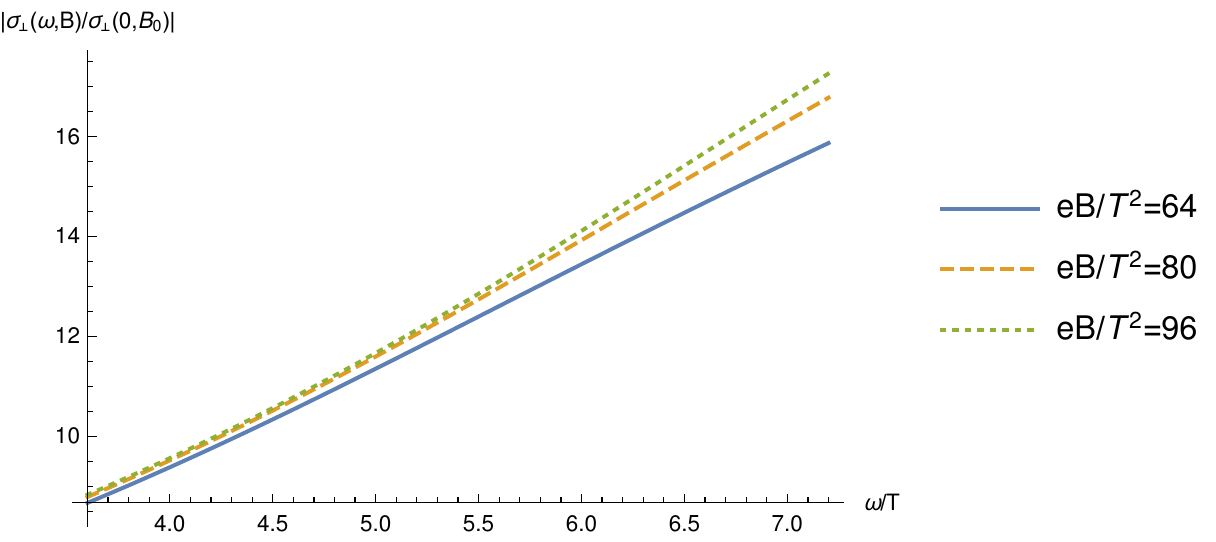}
  \caption{$\o$-dependence of the transverse conductivity for three large values of $B$ with $\t\simeq 0.17\text{fm}$, i.e., $\t T\simeq0.3$. We take $\t_B\simeq0.5\text{fm}-1\text{fm}$ corresponding to $\o/T\simeq3.6-7.2$ so that $\o\lesssim\sqrt{eB}$.  The conductivity is normalized by its value at $eB_0/T^2=1.6$. An enhancement of conductivity is indeed observed.}
  \label{fig:o_dpd2}
\end{figure}

\begin{figure}
  \includegraphics[width=0.6\textwidth]{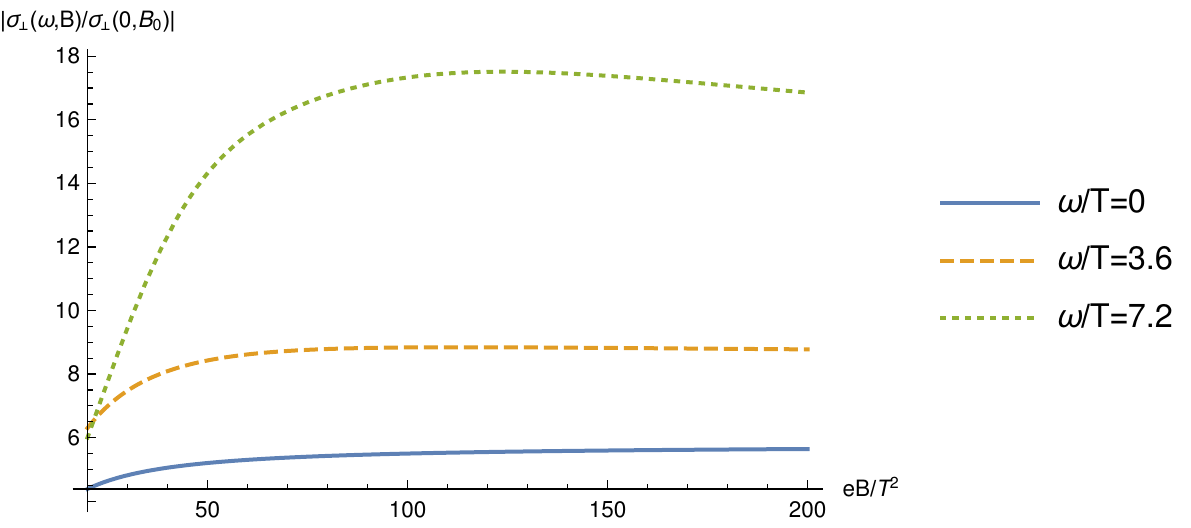}
  \caption{$B$-dependence of the transverse conductivity for three different values of $\o$ with $\t=0.17\text{fm}$, i.e., $\t T\simeq0.3$. We normalize the conductivity by its value at $eB_0/T^2=1.6$. The conductivity approaches a larger value for a higher $\o$. The approaching behavior is from below in the solid line for $\o/T=0$ or from above for high frequencies.}
  \label{fig:b_dpd}
\end{figure}

Let's estimate the parameters for the phenomenology of QGP produced in heavy ion collisions. We take the peak value of magnetic field created in off-central heavy ion collisions as $eB\simeq m_\pi^2-10m_\pi^2$. Since the magnetic field is most relevant at early stage of QGP evolution, we use initial temperature of QGP to set $T=350\text{MeV}$. We assume a neutral QGP with $\m_q=0$, i.e. carrying no net quark density and electric charge density. While we do not have a realistic estimate of relaxation time in strong magnetic field, we use the lattice result of conductivity to estimate the corresponding relaxation time $\t_0$ as \cite{Gorbar:2016qfh}
\begin{align}
  \s=\frac{e^2c^2\t_0}{3}\c.
\end{align}
Taking the lattice result for conductivity $\s/T\simeq 0.37\sum_fe^2Q_f^2$ from \cite{Ding:2010ga}, see also \cite{Gupta:2003zh,Aarts:2007wj,Ding:2016hua} and the free theory result for the susceptibility $\c=\frac{N_cN_fT^2}{3}$. For two flavor case, we obtain $\t_0T\simeq 0.3$. We assume the relaxation time in strong magnetic field is not changed significantly $\t T\simeq 0.3$. For the characteristic frequency of the electric field, we use the life time of electromagnetic field $\t_B$ as a proxy giving $\frac{\o}{T}=\frac{2\pi}{\t_BT}\simeq 3.6-17.9$ with $\t_B\simeq 0.2\text{fm}-1\text{fm}$. Comparing with other dimensionless parameters $\t T=0.3$ and $eB/T^2\simeq 0.16-1.6$, we see the LLL approximation cannot be strictly justified. A definitive answer is not possible without including contribution from higher Landau levels (HLL), which we leave for future studies. We proceed to plot the $\o$ and $B$ dependence of $|\s_{\perp}|$ in Fig.~\ref{fig:o_dpd1} through Fig.~\ref{fig:b_dpd}. 


\section{Summary and Outlook}\label{sec:summary}

In summary, we have formulated a chiral kinetic theory with Landau level basis. It extends the existing CKT framework valid for $B\ll \bp^2$ to the regime of arbitrary magnetic field with $B\sim \bp^2$, for which summing over all Landau level is needed. The theory can be significantly simplified in strong magnetic field, where LLL approximation can be used. The theory also assumes weak electric field and spacetime gradient as in conventional CKT, i.e. $E,\pd_X\ll \bp^2$.

As an application, we use the new chiral kinetic theory to calculate the transverse conductivity of a magnetized QED and QCD plasma. For simplicity, we consider strong magnetic field limit where LLL approximation can be justified and we introduce dissipation by relaxation time approximation. We find a qualitative different behavior from the longitudinal conductivity: while the longitudinal conductivity grows linearly with the magnetic field and is proportional to the relaxation time, the transverse conductivity is inversely proportional to the relaxation time. In addition, we find the transverse conductivity as a function of frequency from response to time-dependent electric field. An enhancement of the conductivity is observed, which may be relevant for the dynamics of magnetic field in heavy ion collisions.

The chiral kinetic theory we discuss has a preferred frame, the frame with constant magnetic field (magnetic frame for short). In the study of conductivity, this is also the plasma frame. It would be interesting to discuss the covariance of the theory under Lorentz transformation. On phenomenological side, it is also interesting to study transport phenomena where the plasma frame do not coincide with the magnetic frame.

The present theory still lacks a collisonal term, which is responsible for the dissipation of the system. In \cite{Fukushima:2019ugr,Fukushima:2017lvb} collisional terms between all LL states have been calculated. This allows us to calculate longitudinal conductivity, since the system remains in the space spanned by the LL basis under longitudinal electric field. Unfortunately this would not be true for the case with transverse electric field. It would be interesting to find out the explicit basis and the corresponding collisional term. We leave it for future work.

\begin{acknowledgments}
We are grateful to Jianhua Gao, Koichi Hattori, Yoshimasa Hidaka, Defu Hou, Xu-Guang Huang, Jinfeng Liao, Shi Pu, Qun Wang and Pengfei Zhuang for useful discussions. We also thank the 13th workshop on QCD phase transition and relativistic heavy-ion physics for providing an stimulating environment in the final stage of this work. S.L. is in part supported by NSFC under Grant Nos 11675274 and 11735007.
\end{acknowledgments}

\appendix

\section{LL Wigner function as lowest order solution}

The Wigner function for system of fermions in constant magnetic field has been worked out in \cite{Sheng:2017lfu,Gorbar:2017awz,Sheng:2018jwf}. Specializing to the case of our interest $m=0$ for right-handed fermions, the Wigner function reduces to the following
\begin{align}\label{Wigner_LL}
  &  \tW^{(0)}=f_+(p_0)\d(p_0-E_{p_z})W_+^{(0)}({\bp})+f_-(p_0)\d(p_0+E_{p_z})W_-^{(0)}(\bp),\no
  &  \tW^{(n)}=f_+(p_0)\d(p_0-E_{p_z})W_{+,s}^{(n)}(\bp)+f_-(p_0)\d(p_0+E_{p_z})W_{-,s}^{(n)}(\bp).
\end{align}
for LLL and HLL states respectively, with $n$ and $s$ being labels for Landau level and spin. Each case is a sum over positive/negative energy Landau level states indicated by the subscript $r=\pm$. The structure function $W_{r,s}^{(n)}$ can be found in appendix of \cite{Sheng:2017lfu} as
\begin{align}\label{Wigner_structure}
  &W_r^{(0)}=\frac{1}{(2\pi)^3}\frac{2E_{p_z}^{(0)}+rp_z}{2E_{p_z}^{(0)}}\mathrm{exp}\(-\frac{p_T^2}{B}\)
      \begin{pmatrix}
        1& 0\\
        0& 0
      \end{pmatrix}, \no
      &W_{r,s}^{(n)}=\frac{1}{(2\pi)^3}\frac{2E_{p_z}^{({n})}+rs\sqrt{p_z^2+2nB}}{2E_{p_z}^{(n)}}\mathrm{exp}\(-\frac{p_T^2}{B}\)
      \begin{pmatrix}
        c_n^2I_{nn}& c_nd_{n-1}I_{n,n-1}\\
        c_nd_{n-1}I_{n-1,n}& d_{n-1}^2I_{n-1,n-1}
      \end{pmatrix}.
\end{align}
The entries of matrix elements can be found in appendix of \cite{Sheng:2017lfu}, from which we can extract the components of Wigner function $F$ and $j_i$:
\begin{align}\label{comp_exp}
  F_r^{(0)}=j_{3r}^{(0)}=\frac{1}{(2\pi)^3}f_r(p_0)\frac{E_{p_z}^{{(0)}}+rp_z}{2E_{p_z}^{{(0)}}}\L^{(0)}(p_T), \no
  F_{rs}^{(n)}=\frac{1}{(2\pi)^3}f_r(p_0)\frac{E_{p_z}^{{(n)}}+rs\sqrt{p_z^2+2nB}}{2E_{p_z}^{{(n)}}}\(\L_+^{(n)}(p_T)+\frac{sp_z}{\sqrt{p_z^2+2nB}}\L_-^{(n)}(p_T)\), \no
  j_{3rs}^{(n)}=\frac{1}{(2\pi)^3}f_r(p_0)\frac{E_{p_z}^{{(n)}}+rs\sqrt{p_z^2+2nB}}{2E_{p_z}^{{(n)}}}\(\frac{sp_z}{\sqrt{p_z^2+2nB}}\L_+^{(n)}(p_T)+\L_-^{(n)}(p_T)\), \no
  j_{irs}^{(n)}=\frac{1}{(2\pi)^3}f_r(p_0)\frac{E_{p_z}^{{(n)}}+rs\sqrt{p_z^2+2nB}}{2E_{p_z}^{{(n)}}}\(\frac{2nB}{\sqrt{p_z^2+2nB}}\frac{sp_i}{p_T^2}\L_+^{(n)}(p_T)\),
\end{align}
with $i={1,2}$ labeling the transverse directions. 
The functions $\L_\pm$ are defined by
\begin{align}\label{Lambda}
  \L^{(0)}(p_T)=2\mathrm{exp}\(-\frac{p_T^2}{B}\),
  \L_\pm^{(n)}(p_T)=(-1)^n\big[L_n(\frac{2p_T^2}{B})\mp L_{n-1}(\frac{2p_T^2}{B})\big]\mathrm{exp}\(-\frac{p_T^2}{B}\).
\end{align}
Let's discuss some key properties of the above contribution to Wigner function from individual Landau level. The LLL states has a linear dispersion relation $E_{p_z}^{{(0)}}={|p_z|}$\footnote{We don't shift energy by the chemical potential $p_0\to p_0+\m+\m_5$ as is done in \cite{Sheng:2017lfu} since $\m$ and $\m_5$ can be shifted into distribution function $f_r$.}. Moreover the factor $\frac{E_{p_z}^{(0)}+rp_z}{2E_{p_z}^{(0)}}$ indicates that for nonvanishing contribution $p_z=rE_{p_z}=p_0$. In other words, for positive(negative) energy states $p_z>0(p_z<0)$. There is only one spin that aligns with the magnetic field for LLL state. For HLL states, there are two possible spin alignment $s=\pm1$. The dispersion is $E_{p_z}^{{(n)}}=\sqrt{p_z^2+2nB}$. The factor $\frac{E_{p_z}^{{(n)}}+rs\sqrt{p_z^2+2nB}}{2E_{p_z}^{{(n)}}}$ indicates $rs=1$, i.e. positive(negative) energy states has spin alignment(anti-alignment), which is expected from positive helicity of right-handed fermions.
Now we are ready to verify the contribution from individual Landau level does satisfy the EOM \eqref{eom_comp}. We look at LLL first. The nontrivial constraint equations are given by
\begin{align}
  &p_0F_r^{(0)}-p_zj_{3r}^{(0)}=0, \no
  &{-p_0j_{3r}^{(0)}+p_zF_r^{(0)}=0}.
\end{align}
They are satisfied by the property $p_0=p_z$ discussed above. The nontrivial transport equations are given by
\begin{align}
  \D_iF_r^{(0)}-2\e^{ij}{p_j}j_{3r}^{(0)}=0, 
\end{align}
with $\e^{ij}$ being Levi-Civita symbol in transverse directions. They are satisfied by property of $\L^{(0)}(p_T)$. The HLL is a little complicated with the following constraint equations:
\begin{align}\label{hll_cons}
  &p_0F_{rs}^{(n)}-p_ij_{irs}^{(n)}-p_zj_{3rs}^{(n)}=0,\no
  &-p_0j_{irs}^{(n)}+p_iF_{rs}^{(n)}+\frac{1}{2}\e^{ij}\D_jj_{3rs}^{(n)}=0, \no
  &-p_0j_{3rs}^{(n)}+p_zF_{rs}^{(n)}+\frac{1}{2}\e^{ij}\D_ij_{jrs}^{(n)}=0.
\end{align}
The first equation of \eqref{hll_cons} is satisfied by noting $r=s$ and using the dispersion $p_0=r\sqrt{p_z^2+2nB}$. The second and third equations can be shown to hold by the following properties
\begin{align}\label{Lam_prop}
  &\L_-^{(n)}+\frac{\pd}{\pd \(p_T^2/B\)}\L_+^{(n)}=0,\no
  &\(-\frac{2n}{p_T^2/B}+1\)\L_+^{(n)}+\frac{\pd}{\pd \(p_T^2/B\)}\L_-^{(n)}=0.
\end{align}
The second equation of \eqref{hll_cons} is satisfied as
\begin{align}
&-p_0j_{irs}^{(n)}+p_iF_{rs}^{(n)}+\frac{1}{2}\e^{ij}\D_jj_{3rs}^{(n)} \no 
\sim&-p_0\(\frac{2nB}{\sqrt{p_z^2+2nB}}\frac{sp_i}{p_T^2}\L_+^{(n)}(p_T)\)+p_i\(\L_+^{(n)}(p_T)+\frac{sp_z}{\sqrt{p_z^2+2nB}}\L_-^{(n)}(p_T)\)\no
&+\frac{B}{2}\frac{\pd}{\pd p_i}\(\frac{sp_z}{\sqrt{p_z^2+2nB}}\L_+^{(n)}(p_T)+\L_-^{(n)}(p_T)\)\no
\sim&\[\(-\frac{2n}{p_T^2/B}+1\)\L_+^{(n)}+\frac{\pd}{\pd \(p_T^2/B\)}\L_-^{(n)}\]+\frac{sp_z}{\sqrt{p_z^2+2nB}}\[\L_-^{(n)}+\frac{\pd}{\pd \(p_T^2/B\)}\L_+^{(n)}\]=0.
\end{align}
The third equation of \eqref{hll_cons} vanishes similarly
\begin{align}
&-p_0j_{3rs}^{(n)}+p_zF_{rs}^{(n)}+\frac{1}{2}\e^{ij}\D_ij_{jrs}^{(n)}\no
\sim&-p_0\(\frac{sp_z}{\sqrt{p_z^2+2nB}}\L_+^{(n)}(p_T)+\L_-^{(n)}(p_T)\)+p_z\(\L_+^{(n)}(p_T)+\frac{sp_z}{\sqrt{p_z^2+2nB}}\L_-^{(n)}(p_T)\)\no
&-\frac{B}{2}\[\frac{\pd}{\pd p_i}\(\frac{2nB}{\sqrt{p_z^2+2nB}}\frac{sp_i}{p_T^2}\L_+^{(n)}(p_T)\)\]\no
\sim&-\[\L_-^{(n)}+\frac{\pd}{\pd \(p_T^2/B\)}\L_+^{(n)}\]=0.
\end{align}
It remains to show the HLL contribution to Wigner function also satisfies the transport equations. The time and longitudinal components of transport equations simply vanish by property of $j_{jrs}^{(n)}$.
\begin{align}
  &\D_0F_{rs}^{(n)}+\D_3j_{3rs}^{(n)}+\D_ij_{irs}^{(n)}\sim \e^{ij}\frac{\pd}{\pd p_i}j_{jrs}^{(n)}=0, \no
  &\D_0j_{3rs}^{(n)}+\D_3F_{rs}^{(n)}-2\e^{ij}p_ij_{jrs}^{(n)}\sim -\e^{ij}p_ij_{jrs}^{(n)}=0.
\end{align}
The transverse components of transport equations read
\begin{align}
  &\D_0j_{irs}^{(n)}+\D_iF_{rs}^{(n)}-2\e^{ij}p_jj_{3rs}^{(n)}+2\e^{ij}p_zj_{jrs}^{(n)} \no
  \sim &-2\e^{ij}p_j\frac{\pd}{\pd (p_T^2/B)}\(\L_+^{(n)}+\frac{sp_z}{\sqrt{p_z^2+2nB}}\L_-^{(n)}\)-2\e^{ij}p_j\(\frac{sp_z}{\sqrt{p_z^2+2nB}}\L_+^{(n)}+\L_-^{(n)}\) \no
  &+2\e^{ij}p_z\(\frac{2nB}{\sqrt{p_z^2+2nB}}\frac{sp_j}{p_T^2}\L_+^{(n)}\).
\end{align}
It also vanishes by the properties in \eqref{Lam_prop}.

As a last comment, we emphasize that the EOM are satisfied by the momentum distribution of individual LL contribution independent of the energy distribution, i.e. the distribution function $f_\pm(p_0)$ in \eqref{Wigner_LL} is arbitrary. This is so because the system under consideration is free without interaction between LL states.

\section{self-consistency of the over-determined system}

For clearance, we reproduce the over-determined system \eqref{eom_comp} below
\begin{align}\label{eom2}
  \D_0F+\D_ij_i=0, \no
  \D_0j_i+\D_iF-2\e^{ijk}p_jj_k=0 \no
  p_0F-p_ij_i=0, \no
  -p_0j_i+p_iF+\frac{1}{2}\e^{ijk}\D_jj_k=0.
\end{align}
We wish to show our claim: once the constraint equations in \eqref{eom2} are satisfied by initial condition, it continues to hold as the system evolves with the dynamical equations in \eqref{eom2}. For simplicity we assume the system is subject to homogeneous electromagnetic field. Without loss of generality, we assume the constraint equations are satisfied at $t=0$. The self-consistency of the system requires the time derivative of the constraint equations vanish at $t=0$.
\begin{align}\label{initial}
  &\pd_t\(p_0F-p_ij_i\)=0, \no
  &\pd_t\(p_0j_i-p_iF-\frac{1}{2}\e^{ijk}\D_jj_k\)=0.
\end{align}
Note that this is a necessary condition but not a sufficient one. A sufficient condition would be to show any order time derivatives vanish at $t=0$. We shall not attempt it here.
To see that \eqref{initial} is indeed true, we use $\pd_t=\D_0-F_{0i}\frac{\pd}{\pd p_i}$. On the time slice $t=0$, the term $F_{0i}\frac{\pd}{\pd p_i}$ contains no time derivative, thus is fully determined by the initial condition. It obviously vanishes and thus can be dropped. Using $\D_\m{P_\n}={P_\n}\D_\m{-}F_{\m\n}$, or explicitly $\D_\m p_0=p_0\D_\m-F_{\m0}$ and $\D_\m p_i=p_i\D_\m+F_{\m i}$, we can simplify the first equation of \eqref{initial}
\begin{align}
  \pd_t\(p_0F-p_ij_i\)=\D_0\(p_0F-p_ij_i\)
  =p_0\D_0F-p_i\D_0j_i-F_{0i}j_i.
\end{align}
Using the dynamical equations, we obtain
\begin{align}
  &p_0\D_0F-p_i\D_0j_i-F_{0i}j_i=-p_0\D_ij_i-F_{0i}j_i-p_i\(-\D_iF+2\e^{ijk}p_jj_k\) \no
  &=-\D_i\(p_0j_i-p_iF\)=-\D_i\(p_0j_i-p_iF-\frac{1}{2}\e^{ijk}\D_jj_k\)=0,
\end{align}
where in the last equality we have used the initial condition.
The second equation of \eqref{initial} requires a little more work.
\begin{align}
  &\pd_t\(-p_0j_i+p_iF+\frac{1}{2}\e^{ijk}\D_jj_k\)=\D_0\(-p_0j_i+p_iF+\frac{1}{2}\e^{ijk}\D_jj_k\) \no
  =&-p_0\D_0j_i+F_{0i}F+p_i\D_0F+\frac{1}{2}\e^{ijk}\D_j\D_0j_k.
\end{align}
Using the dynamical equations, we obtain
\begin{align}
  &-p_0\D_0j_i+F_{0i}F+p_i\D_0F+\frac{1}{2}\e^{ijk}\D_j\D_0j_k \no
  =&\frac{1}{2}\e^{ijk}\D_j\(-\D_kF+2\e^{kmn}p_mj_n\)+p_i\(-\D_jj_j\)-p_0\(-\D_iF+2\e^{ijk}p_jj_k\)+F_{0i}F \no
  =&\D_j\(p_ij_j\)-\D_j\(p_jj_i\)-p_i\D_jj_j+\D_i\(p_jj_j\)-2\e^{ijk}p_0p_jj_k \no
  =&-p_j\(\D_jj_i-\D_ij_j\)-2\e^{ijk}p_0p_jj_k.
\end{align}
To see the last expression vanishes, we use initial condition again:
\begin{align}
  0=2\e^{imn}p_m\(-p_0j_n+p_nF+\frac{1}{2}\e^{njk}\D_jj_k\)=-\(p_j{\D_jj_i}-p_j{\D_ij_j}\)-2\e^{ijk}p_0p_jj_k.
\end{align}
Therefore, we have shown the self-consistency of the over-determined system \eqref{eom2}.

\section{Momenta Integration}

In this appendix, we collect details in doing the momenta integrals to obtain the induced current density. To be specific, we point the electric field along $x$-direction. It leads to the following Wigner function components
\begin{align}\label{FdF}
&\d j_1=E\frac{4p_z\t F-\(2B\t+\frac{1}{\t}\)\frac{\pd F}{\pd p_0}}{\(2B\t+\frac{1}{\t}\)^2+4p_z^2},\no
&\d j_2=-E\frac{2p_z\frac{\pd F}{\pd p_0}+2\(2B\t+\frac{1}{\t}\)\t F}{\(2B\t+\frac{1}{\t}\)^2+4p_z^2}.
\end{align}
We first note that the ${\bp}_T$ dependence is proportional to $\mathrm{exp}\(-\frac{p_T^2}{B}\)$, thus the integration over transverse momenta is easy
\begin{align}
\int d^2\bp_T\,\mathrm{exp}\(-\frac{p_T^2}{B}\)=\p B.
\end{align}
As we have shown in \eqref{pic_trans}, $F$ represented by positively and negatively charged LLL states are given by
\begin{align}\label{F_pm}
  F=\frac{2}{(2\pi)^3}\mathrm{exp}\(-\frac{p_T^2}{B}\)\(\frac{1}{e^{\b(p_0-\m)}+1}\th(p_0)-\frac{1}{e^{\b(-p_0+\m)}+1}\th(-p_0)\)\d(p_0-p_z),
\end{align}
where we have dropped the $1$ from vacuum contribution. To obtain the integration of $\d j_1$ and $\d j_2$ in \eqref{FdF}, we separate $\d j_i$ into $\d j_{i,F}\propto F$ and $\d j_{i,D}\propto\frac{\pd F}{\pd p_0}$. The integration of the former is straight forward. For the latter, we need to be careful as the integrands are not differentiable at $p_0=0$, where we need to split the integration interval:
\begin{align}
&\int dp_0dp_z\frac{\pd\d(p_0-p_z)}{\pd p_0}\(\frac{2B\t+\frac{1}{\t}}{\(2B\t+\frac{1}{\t}\)^2+4p_z^2}\sum_{r=\pm}\frac{r\th(rp_0)}{e^{r\b(p_0-\m)}+1}\)\no
=&-\int dp_0dp_z\frac{\pd\d(p_0-p_z)}{\pd p_z}\(\frac{2B\t+\frac{1}{\t}}{\(2B\t+\frac{1}{\t}\)^2+4p_z^2}\sum_{r=\pm}\frac{r\th(rp_0)}{e^{r\b(p_0-\m)}+1}\)\no
=&\lim_{\e\to0}\int dp_0\(\int_{-\infty}^{-\e}+\int_{\e}^{\infty}\)dp_z\d(p_0-p_z)\frac{\pd}{\pd p_z}\(\frac{2B\t+\frac{1}{\t}}{\(2B\t+\frac{1}{\t}\)^2+4p_z^2}\sum_{r=\pm}\frac{r\th(rp_0)}{e^{r\b(p_0-\m)}+1}\)\no
=&\lim_{\e\to0}\(\int_{-\infty}^{-\e}+\int_{\e}^{\infty}\)dp_z\frac{\pd}{\pd p_z}\(\frac{2B\t+\frac{1}{\t}}{\(2B\t+\frac{1}{\t}\)^2+4p_z^2}\sum_{r=\pm}\frac{r\th(rp_z)}{e^{r\b(p_z-\m)}+1}\)\no
=&\lim_{\e\to0}\frac{2B\t+\frac{1}{\t}}{\(2B\t+\frac{1}{\t}\)^2+2\e^2}\(\sum_{r=\pm}\frac{r\th(-r\e)}{e^{r\b(-\e-\m)}+1}-\sum_{r=\pm}\frac{r\th(r\e)}{e^{r\b(\e-\m)}+1}\)=-\frac{1}{2B\t+\frac{1}{\t}}
\end{align}
Similarly,
\begin{align}
&\int dp_0dp_z\frac{\pd\d(p_0-p_z)}{\pd p_0}\(\frac{\(2B\t+\frac{1}{\t}\)p_z}{\(2B\t+\frac{1}{\t}\)^2+4p_z^2}\sum_{r=\pm}\frac{r\th(rp_0)}{e^{r\b(p_0-\m)}+1}\)\no
=&\lim_{\e\to0}\frac{\(2B\t+\frac{1}{\t}\)\e}{\(2B\t+\frac{1}{\t}\)^2+2\e^2}\(\sum_{r=\pm}\frac{-r\th(-r\e)}{e^{r\b(-\e-\m)}+1}-\sum_{r=\pm}\frac{r\th(r\e)}{e^{r\b(\e-\m)}+1}\)=0
\end{align} 
Using the above identities, we get
\begin{align}
\d J_{1,F}&=\int d^4P\d j_{1,F}\no
&=8E\int\frac{d^3\bp}{(2\p)^3}\frac{p_z\t}{\(2B\t+\frac{1}{\t}\)^2+4p_z^2}\mathrm{exp}\(-\frac{p_T^2}{B}\)\sum_{r=\pm}\frac{r\th(rp_z)}{e^{r\b(p_z-\m)}+1}\no
&=\frac{4BE}{(2\p)^2}\int dp_z\frac{p_z\t}{\(2B\t+\frac{1}{\t}\)^2+4p_z^2}\sum_{r=\pm}\frac{r\th(rp_z)}{e^{r\b(p_z-\m)}+1}\no
&=\frac{4BE}{(2\p)^2}\(\int_{0}^{\infty}\frac{p_z}{e^{\b(p_z-\m)}+1}+\int_{0}^{\infty}\frac{p_z}{e^{\b(p_z+\m)}+1}\)\frac{\t}{\(2B\t+\frac{1}{\t}\)^2+4p_z^2}dp_z\no
&\xrightarrow{B\gg p_z^2}\frac{4BE}{(2\p)^2}\int_{0}^{\infty}\(\frac{1}{e^{\b(p_z+\m)}+1}+\frac{1}{e^{\b(p_z-\m)}+1}\)p_z\frac{1}{4B^2\t}dp_z=\frac{3\m^2+\p^2T^2}{24\p^2B\t}E
\end{align}
\begin{align}
\d J_{1,D}&=\int d^4P\d j_{1,D}\no
&=-2E\int\frac{d^3\bp}{(2\p)^3}\mathrm{exp}\(-\frac{p_T^2}{B}\)\frac{\pd}{\pd p_z}\(\frac{2B\t+\frac{1}{\t}}{\(2B\t+\frac{1}{\t}\)^2+4p_z^2}\sum_{r=\pm}\frac{r\th(rp_z)}{e^{r\b(p_z-\m)}+1}\)\no
&=-\frac{BE}{(2\p)^2}\int dp_z\frac{\pd}{\pd p_z}\(\frac{2B\t+\frac{1}{\t}}{\(2B\t+\frac{1}{\t}\)^2+4p_z^2}\sum_{r=\pm}\frac{r\th(rp_z)}{e^{r\b(p_z+\m)}+1}\)\no
&=\frac{BE}{(2\p)^2}\frac{1}{2B\t+\frac{1}{\t}}\xrightarrow{B\gg p_z^2}\frac{E}{8\p^2\t}\(1-\frac{1}{2B\t^2}\)
\end{align}
\begin{align}
\d J_{2,F}&=\int d^4P\d j_{2,F}\no
&=-4E\int\frac{d^3\bp}{(2\p)^3}\frac{2B\t^2+1}{\(2B\t+\frac{1}{\t}\)^2+4p_z^2}\mathrm{exp}\(-\frac{p_T^2}{B}\)\sum_{r=\pm}\frac{r\th(rp_z)}{e^{r\b(p_z-\m)}+1}\no
&=-\frac{2BE}{(2\p)^2}\int dp_z\frac{2B\t^2+1}{\(2B\t+\frac{1}{\t}\)^2+4p_z^2}\sum_{r=\pm}\frac{r\th(rp_z)}{e^{r\b(p_z-\m)}+1}\no
&=-\frac{2BE}{(2\p)^2}\(\int_{0}^{\infty}\frac{1}{e^{\b(p_z-\m)}+1}+\int_{0}^{\infty}\frac{-1}{e^{\b(p_z+\m)}+1}\)\frac{2B\t^2+1}{\(2B\t+\frac{1}{\t}\)^2+4p_z^2}dp_z\no
&\xrightarrow{B\gg p_z^2}-\frac{2BE}{(2\p)^2}\int_{0}^{\infty}\(\frac{1}{e^{\b(p_z-\m)}+1}-\frac{1}{e^{\b(p_z+\m)}+1}\)\frac{1}{2B}dp_z\no
&=-\frac{\m BE}{4\p^2B}
\end{align}
\begin{align}
\d J_{2,D}&=\int d^4P\d j_{2,D}=0
\end{align}
Here we've used the relevant integrals,
\begin{align}
&\int_{0}^{\infty}\(\frac{1}{e^{\b(p_z-\m)}+1}-\frac{1}{e^{\b(p_z+\m)}+1}\)dp_z=\m,\no
&\int_{0}^{\infty}\(\frac{1}{e^{\b(p_z-\m)}+1}+\frac{1}{e^{\b(p_z+\m)}+1}\)p_zdp_z=\frac{\m^2}{2}+\frac{\p^2T^2}{6}
\end{align}
Finally, using $\d J_i=\d J_{i,F}+\d J_{i,D}$ we obtain the transverse currents.


\end{document}